\documentclass[12pt]{amsart}

\newcommand{\R}{\mathbb R}
\newcommand{\x}{\mathbf x}
\newcommand{\e}{\varepsilon}
\newcommand{\CC}{\mathbb C}
\newcommand{\Sp}{{Sp}}
\newcommand{\GL}{{GL}}
\newcommand{\Sc}{\mathcal S}

\newcommand{\E}{\mathcal E}

\newcommand{\SG}{\mathcal{SG}}
\newcommand{\GF}{\mathcal{GF}}

\newcommand{\Stab}{\mathop{\rm Stab}\nolimits}

\newcommand{\LGr}{LGr}

\newcommand{\Tr}{\mathop{\rm Tr}\nolimits}

\begin{document}
\title[Infinite dimensional analog of the Weil representation]
{Infinite dimensional analog of the Weil representation in the space of distributions}
\author{A. V. Stoyanovsky}
\email{alexander.stoyanovsky@gmail.com}
\address{Independent Moscow University}
\begin{abstract}
We construct a new version of infinite Grassmannian and infinite dimensional analog
of the Weil representation of the affine symplectic group in the space of distributions.
We give definition of a mathematical solution of the quantum field theory Schr\"odinger
equation in the constructed space, and give examples of solutions of this equation.
\end{abstract}
\maketitle

\section*{Introduction}

Infinite Grassmannian [1] and the Shale--Berezin representation [2, 3] of infinite dimensional
symplectic group in the Fock space, play an important role in mathematical constructions
of two dimensional quantum field theory. Two dimensions are essential here:
most of constructions of two dimensional theory cannot be generalized to greater dimensions.
One can say that the Hilbert Fock space is ``wide enough'' for giving mathematical sense
to equations of two dimensional quantum field theory and for solving these equations,
while for multidimensional case it is not wide enough. A way to overcome this difficulty
proposed in the present paper, is to extend the Fock space. This is similar to the fact that
for solving the wave equation in two dimensions, it suffices to make computations with
usual functions (due to separation of variables), while in greater dimensions,
usual functions are insufficient, and one should deal with distributions.

The purpose of this paper is to construct an infinite dimensional analog of
the space of distributions, to give definition of a solution of the quantum field theory
Schr\"odinger equation and its relativistically invariant generalization
in the constructed space, and to give examples of solutions.

The idea of our approach is stated in the papers [4], [6] (the results of [4] are announced in [5]),
but has not been fully realized in these papers. This idea is to generalize to infinite dimensional case
the Weil representation of the affine symplectic group in the space of distributions.
For such generalization, we use the geometric realization of the Weil representation in certain
space of functions on the complex Lagrangian Grassmannian. This space of functions is explicitly
described in [4]. The passage from the usual realization of the Weil representation
to the geometric realization is given by integral transform with Gaussian kernel. In other words,
in the geometric realization points of the Lagrangian Grassmannian correspond to Gaussian functions
in the usual realization, which are transformed under the action of the symplectic group
in the most simple way.

For generalization to infinite dimensional case, one should first of all construct infinite
dimensional generalization of the (Lagrangian) Grassmannian. This generalization
was sketched at the end of [4].
The usual generalization of the Grassmannian [1] does not suit us, because it leads to the
unitary representation in the Fock space, which is the analog of the Weil representation in the space $L_2$
but not in the space of distributions. We shall modify the definition of the Grassmannian and the symplectic group
from [1]. Recall that in [1, 2, 3], the restricted general linear group of a Hilbert space $H=H_+\oplus H_-$
is defined as the group of $2\times2$-matrices of operators
$\left(\begin{array}{cc}A&B\\C&D\end{array}\right)$, where $B$ and $C$ are Hilbert--Schmidt operators.
We modify this condition requiring $B$ to be a trace class operator, and $C$ to be an arbitrary operator.
With such modification the main constructions from [1], for instance, the central extension,
the Grassmannian and the determinant line bundle, nevertheless hold; for example, the cocycle
$\Tr(b_1c_2-b_2c_1)$ on the Lie algebra is still well defined. It will be convenient to change the
construction a little more. Instead of Hilbert space $H_+$, consider a nuclear space $V$
(for example, the Schwartz space),
instead of $H_-$ consider the dual space $V'$,
and instead of the restricted general linear group, consider simply the general linear group of the space
$V\oplus V'$. This group is obtained from the modified group described above by going to the standard
limit from a Hilbert space to a countably Hilbert nuclear space [7], and the main constructions from [1]
again hold for this group. The Grassmannian is defined as the orbit of the subspace $V\subset V\oplus V'$
under the action of the general linear group.

The rest of constructions and generalizations are more or less straightforward.
For defining solutions of functional differential equations one uses the idea from [8] on
regularization of a differential operator. An example of solution of the quantum field theory
Schr\"odinger equation is the standard quantization of free scalar field in the Fock space.
Note that our notion of solution of the Schr\"odinger equation agrees, in the principal order
of quasiclassical approximation, with the Maslov--Shvedov method of complex germ in field theory ([9], see also
a brief exposition in [10]).

The paper is organized as follows.  In \S1 we define the integral transform with the Gaussian kernel,
which we call the Gauss--Fourier transform; we show, following [6], its naturalness from the viewpoint
of theory of partial differential equations and compatibility with the method of complex germ;
and we describe the image of certain function spaces under this transform, i.~e.
the geometric realization of the Weil representation. In \S2 we construct generalization of the
geometric realization to infinite dimensional case. Finally, in \S3 we give definition and
examples of solutions of functional differential equations in the spaces constructed in \S2.

The author is deeply grateful to V. P. Maslov for constant support and attention.

\section{The Gauss--Fourier transform}

\subsection{Motivation and definition of the Gauss--Fourier transform}

The most powerful modern methods of solving linear partial differential equations
are based on far developed deep relation between the notions of classical and quantum mechanics.
The role of particle in quantum mechanics is played by a so called wave packet, i.~e.
a rapidly oscillating function localized near a point of the configuration space $\R^n$.
The role of momentum of the particle is played by the direction of oscillation of the wave packet.

In the book [9], V. P. Maslov and O. Yu. Shvedov introduced the deep notion of balanced wave packet,
whose coordinate as well as momentum are localized in a domain with diameter of order $\sqrt h$,
where $h$ is the Planck constant.
The main example of balanced wave packet is a {\it Gaussian wave packet}
\begin{equation}
\psi_{x_0,p_0,Z,S;h}(x)=e^{\frac ih\left(\frac12(x-x_0)^TZ(x-x_0)+p_0(x-x_0)+S\right)}.
\end{equation}
Here $x,x_0,p_0\in\R^n$, $S\in\R$, the symbol $T$ means transposing, and $Z$ is a symmetric complex
$n\times n$-matrix with positive definite imaginary part. Maslov and Shvedov have shown that
under evolution given by the Schr\"odinger equation, the motion of a Gaussian wave packet in
the principal order of quasiclassical approximation (i.~e. up to $o(h)$)
can be described in purely classical terms. Namely, the point $(x_0,p_0)$ of the phase space moves along
a classical trajectory given by the canonical Hamilton equations, the number $S$ behaves as
action along the trajectory, and the matrix $Z$ is transformed under the action of the tangent
symplectic transformation to the Hamiltonian flow along the trajectory: if the tangent symplectic
transformation is given by $\left(\begin{array}{cc}A&B\\C&D\end{array}\right)$, where $A$, $B$, $C$, $D$ are
$n\times n$-matrices, then the matrix $Z$
goes to the matrix $(AZ+B)(CZ+D)^{-1}$. Besides that, one should multiply the wave packet by the
number factor $1/\sqrt{\det(CZ+D)}$.

On the other hand, L. H\"ormander [11] introduced the important notion of wave front of a distribution.
The wave front is a subset of the phase space obtained by pairing (integrating) the distribution with
various wave packets, and considering the asymptotics of this integral as $h\to0$ up to $o(h^N)$
for $N$ arbitrarily large (we shall say shortly, up to $o(h^\infty)$). Roughly speaking,
if the asymptotics is identically 0, then the point of the phase space does not belong to the wave front.

Now if we join together Maslov--Shvedov's and H\"ormander's approaches, then the next natural step would be
to consider the integral
\begin{equation}
\widetilde u(x_0,p_0,Z,S;h)=\int\overline{u(x)}\psi_{x_0,p_0,Z,S;h}(x)dx
\end{equation}
for a given distribution $u(x)$, and considering the asymptotics of this integral up to $o(h^\infty)$. While
the wave front gives only information on singularities of the distribution, this latter
asymptotics would give more complete information. And indeed, one can show that it is possible to {\it completely
reconstruct} the distribution $u(x)$ from this asymptotics.

It is now natural to make one more not big but essential step, and consider the integral (2) for {\it finite} values
of the constant $h>0$, instead of the asymptotics as $h\to0$. Up to a simple change of parameters,
this amounts to the integral transform
\begin{equation}
(\GF u)(Z,p)=\int\overline{u(x)}e^{\frac ih\left(\frac12x^TZx+px\right)}dx,
\end{equation}
where $x\in\R^n$, $p\in\CC^n$, and $Z$ has the same sense as above.
\medskip

{\bf Definition.} We call the integral transform (3) by the {\it Gauss--Fourier transform}.
\medskip

It takes a distribution $u(x)$ (from certain class) to the function $\GF u(Z,p)$.

We believe that the Gauss--Fourier transform (or at least asymptotics of the integral (2) up to $o(h^\infty)$)
should play an essential role in the theory of linear partial differential equations. It strengthens and unifies
H\"ormander's method of wave fronts as well as Maslov--Shvedov's method of complex germ.
Besides that, one can see that some limit cases of the Gauss--Fourier transform amount to the
Fourier transform and to the Radon transform (cf. [4]).

\subsection{The image of certain function spaces under the Gauss--Fourier transform}
The Gauss--Fourier transform is closely related to the projective unitary Weil representation $\rho$
of the affine symplectic group
$\Sp(2n,\R)\widetilde\times\R^{2n}$ in the space of functions of $n$ variables, because the kernel of
this transform
\begin{equation}
\psi_{Z,p}(x)=e^{\frac ih\left(\frac12x^TZx+px\right)}
\end{equation}
is transformed by simple formulas under the action of this group:
\begin{equation}
\rho\left(\begin{array}{cc}A&B\\C&D\end{array}\right)\psi_{Z,p}=\frac{\psi_{(AZ+B)(CZ+D)^{-1},p(CZ+D)^{-1}}\cdot
e^{-\frac i{2h}p(CZ+D)^{-1}Cp^T}}{\sqrt{\det(CZ+D)}},
\end{equation}
\begin{equation}
\begin{aligned}{}
&\rho(p_0)\psi_{Z,p}=\psi_{Z,p+p_0},\\
&(\rho(x_0)\psi_{Z,p})(x)=\psi_{Z,p}(x-x_0)=\psi_{Z,p-x_0^TZ}(x)\cdot e^{\frac ih\left(\frac12x_0^TZx_0-px_0\right)}.
\end{aligned}
\end{equation}
Regarding the Weil representation we refer the reader to the paper [4], and we shall freely use it.

In this Subsection we shall describe the image under the Gauss--Fourier transform of spaces of functions invariant
with respect to the Weil representation. We know five such spaces: the space $L_2$, the Schwartz space $S$,
the dual space $S'$ of tempered distributions, the Gelfand--Shilov space $S_{1/2}^{1/2}$ [12], and the dual space
$(S_{1/2}^{1/2})'$.
We do not know the image of the space $L_2$, and know only conjecturally the images of the two latter spaces.
For description of the image, we need some notations and constructions.

The function $\psi_{Z,p}(x)$ satisfies the equations
\begin{equation}
\left(ih\frac\partial{\partial x_j}+\sum Z_{jk}x_k\right)\psi=-p_j\psi,\ \ j=1,\ldots,n.
\end{equation}
The left hand sides of these equations form a basis of a positive Lagrangian subspace $L$ in the $2n$-dimensional
symplectic complex vector space with the basis $(x_1,\ldots,x_n,ih\partial/\partial x_1$, $\ldots$,
$ih\partial/\partial x_n)$
and with the symplectic form $\omega$ given by commutator of operators. Positivity means that
the Hermitian form
\begin{equation}
\langle v_1,v_2\rangle=\frac 1i\omega(\overline v_1,v_2),\ \ v_1,v_2\in L,
\end{equation}
is positive definite. This yields identification of the set of matrices $Z$ with the open domain in the
complex Lagrangian Grassmannian consisting of positive Lagrangian subspaces and called the Siegel upper
half-plane $\SG$. The group $\Sp(2n,\R)$ acts on the Siegel upper half-plane, which corresponds to the action
$Z\mapsto(AZ+B)(CZ+D)^{-1}$ on matrices $Z$.

The closure $\overline\SG$ of the Siegel upper half-plane consists of the Lagrangian subspaces $L$ for which the form
(8) is nonnegative definite. For each $L\in\overline\SG$ and for each element
$w$ of the dual space $L'$ taking real values on the kernel of the form (8),
the system of equations similar to (7) defines, uniquely up to a number factor, a tempered distribution
$\psi=\psi_{L,w}$. Thus, we obtain the equivariant complex line bundle
$\mu$ on the space $\widetilde\E$ of pairs $(L,w)$, whose fiber over a point $(L,w)$ is $\CC\psi_{L,w}$.
Over the Siegel upper half-plane, i.~e. on the subspace $\E$ of pairs $(L,w)$, $L\in\SG$, $w\in L'$,
the line bundle $\mu$ is trivialized by the functions $\psi_{Z,p}$.
The transition functions of this line bundle corresponding to the action of the affine symplectic group,
are given by formulas (5, 6).
The bundle $\mu$ extends to the closure $\overline\E$ of the manifold $\E$ in the affine Lagrangian Grassmannian, i.~e.
to the set of pairs $(L,w)$, where $L\in\overline\SG$, $w\in L'$. The definition of the fiber of the bundle $\mu$
at a point $(L,w)\in\overline\E$ is the same as above, but the distribution $\psi_{L,w}$ is in this case,
in general, not tempered. It belongs, for instance, to the space $(S_{1/2}^{1/2})'$.

The function $\widetilde u(Z,p)=\GF u$ (3) satisfies the equations
\begin{equation}
\frac\partial{\partial Z_{jk}}\widetilde u = -\frac{ih}2\frac{\partial^2}{\partial p_j\partial p_k}\widetilde u.
\end{equation}
\medskip

{\bf Theorem 1.} {\it Under the Gauss--Fourier transform,

\emph{(i)} the space $S'$ is identified with the space of holomorphic functions on the manifold $\E$ which
satisfy equations \emph{(9)} and have polynomial growth near the boundary $\widetilde\E\setminus\E$;

\emph{(ii)} the Schwartz space $S$ is identified with the space of holomorphic functions on $\E$ which
satisfy equations \emph{(9)} and extend to sections of the bundle $\mu'$ on the space $\widetilde\E$,
continuous together with their derivatives of all orders with respect to the action of the
Lie algebra of the affine symplectic group.
The topology on this space is the topology of uniform convergence on compacts in $\widetilde\E$
together with the derivatives of any order with respect to the action of the
Lie algebra of the affine symplectic group.
}
\medskip

Theorem 1 easily follows from the results of [4].
\medskip

{\bf Conjecture 2.} {\it Under the Gauss--Fourier transform,

\emph{(iii)} the space $(S_{1/2}^{1/2})'$ is identified with the space of all holomorphic functions
on the manifold $\E$ which satisfy equations \emph{(9),} with the topology of uniform convergence
on compacts;

\emph{(iv)} the space $S_{1/2}^{1/2}$ is identified with the space of all holomorphic
functions on the manifold $\E$ which satisfy equations \emph{(9)} and extend to analytical sections
of the bundle $\mu'$ on a neighborhood of the closure $\overline\E$ in the affine Lagrangian Grassmannian,
with the topology of uniform convergence
on compacts in $\overline\E$. }
\medskip

We shall call the four spaces of sections of the bundle $\mu'$ given in Theorem 1 and Conjecture 2 by
{\it geometric realizations} of the Weil representation.

\section{Infinite dimensional generalizations of the geometric realization of the Weil representation}

\subsection{Introduction}
In this Section we construct infinite dimensional generalizations of three out of four geometric
realizations of the Weil representation.
The exception is the geometric realization of the space $S'$ (part (i) of Theorem 1 from \S1), since it is unclear
how to generalize the condition of polynomial growth near the boundary.

Let $V$ be a real nuclear space [7], $V'$ be the dual space.
We must construct the symplectic group $\Sp=\Sp(V\oplus V')$,
the complex Lagrangian Grassmannian $\LGr$ of Lagrangian subspaces $L$ in the complexification
$V_\CC\oplus V'_\CC$, the Siegel upper half-plane $\SG\subset\LGr$, the spaces $\E\subset\widetilde\E$,
the equivariant line bundle $\mu$ on the closure $\overline\E$ in the affine Lagrangian Grassmannian,
and the analog of equations (9), so that in the case $V=\R^n$ these objects would coincide
with the objects from \S1.

The group $\Sp$ is defined in the obvious way.
Regarding topology on it, see [12, 7].
The Grassmannian $\LGr$ is the orbit of the subspace $V_\CC\subset V_\CC\oplus V'_\CC$ under the action
of the group $\Sp(V_\CC\oplus V'_\CC)$. The Siegel upper half-plane, the spaces $\E$, $\widetilde\E$, and $\overline\E$
are defined obviously, using the analog of the form (8). It remains to define the bundle $\mu$ over
$\overline\E$ and the analog of equations (9).

The bundle $\mu$ is obtained from the determinant line bundle on infinite Grassmannian [1].
Let us provide the corresponding constructions.

\subsection{The central extension and the determinant line bundle}

First of all, let us define the central extension $\widetilde\GL_\CC$ of the general linear group
$\GL_\CC=\GL(V_\CC\oplus V'_\CC)$ by the group $\CC^\times=\CC\setminus0$. The group $\widetilde\GL_\CC$
is defined as the quotient group $G/G_1$, where $G$ is the group of pairs $(g,q)$,
$g=\left(\begin{array}{cc}A&B\\C&D\end{array}\right)\in\GL(V_\CC\oplus V'_\CC)$,
$q\in\GL(V_\CC)$, such that
the operator $Aq^{-1}$ has the (Fredholm) determinant [13], and $G_1$ is the subgroup of pairs $(1,q)$, where $q$
has the determinant equal to $1$.

As in [1], one proves that the value of the cocycle of this central extension on two matrices
$\left(\begin{array}{cc}a_1&b_1\\c_1&d_1\end{array}\right)$,
$\left(\begin{array}{cc}a_2&b_2\\c_2&d_2\end{array}\right)$ from the Lie algebra of the group $\GL_\CC$ equals
$\Tr(b_1c_2-b_2c_1)$.

Further, the determinant line bundle over the Grassmannian is obtained from the principal
$\CC^\times$-bundle $\widetilde\GL_\CC/\Stab V_\CC$, where $\Stab V_\CC\subset\GL_\CC$ is the stabilizer
of the subspace $V_\CC\subset V_\CC\oplus V'_\CC$ (over this subgroup the central extension
is trivial, hence we can treat it as a subgroup in $\widetilde\GL_\CC$).

One has the similar bundle over the affine Grassmannian: it is obtained by replacement of the general linear group
with the affine linear group. We shall also call it by the determinant line bundle.

\subsection{The final part of the construction}

The line bundle $\mu$ is the square root of restriction of the determinant line bundle
onto the subspace $\overline\E$. This square root exists and is unique due to the following Conjecture.
\medskip

{\bf Conjecture.} {\it The space $\overline\E$ is contractible.}
\medskip

One has the action on this line bundle of the central extension of the group $\Sp\widetilde\times(V\oplus V')$
by $\CC^\times$.

It remains to define the analog of equations (9). It reads
\begin{equation}
L_b=-\frac{ih}2L_{v_1}L_{v_2},
\end{equation}
where $L_b$ is the Lie derivative of the action of the matrix
$\left(\begin{array}{cc}0&b\\0&0\end{array}\right)$ from the Lie algebra of the group $\Sp$,
$b=v_1v_2$, $v_1,v_2\in V$, and $L_{v_i}$, $i=1,2$, is the Lie derivative of the action of the element
$v_i\in V$ of the Lie subalgebra of shifts $V\oplus V'$ in the affine symplectic Lie algebra. Here we can
neglect the central extension, since its restriction to the Lie subalgebras in question is trivial.

Thus, repeating definitions (ii)--(iv) from Theorem 1 and Conjecture 2 in \S1, we obtain constructions
of three topological vector spaces. Choosing one of these spaces, we shall denote it below by the symbol
$\Sc$.

\medskip
{\bf Definition.} We shall call the dual space to any of the three constructed spaces by the
{\it space of distribution functionals}, and denote it by $\Sc'$.
\medskip

{\it Remark.} The constructions of this paper can be generalized to the fermionic case.
Instead of the Weil representation of the group $\Sp(2n,\R)$, one considers the spinor representation of the
group $SO(2n,\R)$. For its geometric realization on the Grassmannian of maximal isotropic subspaces in $\CC^{2n}$,
see [1, Ch.~12]. Instead of the square root of the determinant line bundle, one uses the Pfaffian line bundle, etc.
For details, see [1].

\section{Solutions of functional differential equations}

\subsection{Examples of functional differential equations}

Functional differential (quantum field theory) Schr\"odinger equation reads
\begin{equation}
ih\frac{\partial\Psi}{\partial t}=\int\left(-\frac{h^2}2\frac{\delta^2}{\delta\varphi(\x)^2}+
\frac12\sum_{j=1}^d\varphi_{x^j}(\x)^2+U(t,\x,\varphi(\x))\right)d\x\Psi,
\end{equation}
where $\Psi=\Psi(t)$ is the unknown complex valued ``functional'' of smooth real function $\varphi(\x)$,
$\x=(x^1$, $\ldots$, $x^d)\in\R^d$, $\varphi_{x^j}=\partial\varphi/\partial x^j$, $U(t,\x,\varphi)$ is a fixed
function, and $\delta/\delta\varphi(\x)$ is the functional (also called variational) derivative.

The relativistically invariant generalization of equation (11) reads
\begin{equation}
\begin{aligned}{}
&x^\mu_{s^k}\frac{\delta\Psi}{\delta x^\mu(s)}+
\varphi_{s^k}\frac{\delta\Psi}{\delta\varphi(s)}=0,\ \ k=1,\ldots,d,\\
&ihD_\mu\frac{\delta\Psi}{\delta x^\mu(s)}=
-\frac{h^2}2\frac{\delta^2\Psi}{\delta\varphi(s)^2}
+D^\mu D_\mu\left(\frac12 d\varphi(s)^2+U(x(s),\varphi(s))\right)\Psi.
\end{aligned}
\end{equation}
Here $\Psi$ is the unknown ``functional'' of smooth functions
$(x(s)$, $\varphi(s))$, $x=(x^\mu)=(t,\x)$, $\mu=0,\ldots,d$, $x^0=t$,
$s=(s^1,\ldots,s^d)\in\R^d$,
$$
D^\mu=(-1)^\mu \frac{\partial(x^0,\ldots,\widehat{x^\mu},\ldots,x^d)}
{\partial(s^1,\ldots,s^d)}
$$
is the Jacobian with the sign $(-1)^\mu$, the hat over a variable means that the variable is omitted;
raising and lowering of the index $\mu$
goes using the Lorentz metric
$$
dx^2=(dx^0)^2-\sum_{j=1}^d(dx^j)^2;
$$
finally, $d\varphi(s)^2$ is the scalar square of the differential $d\varphi(s)$ of the function $\varphi(s)$
on the surface $x=x(s)$ in the Minkowsky space $\R^{1+d}$.

The first $d$ equations of the system (12) mean that the ``functional'' $\Psi$ is invariant with respect to smooth
changes of variables $s$.

Let us mention one more example of functional differential equation, the {\it quantum Plato problem}:
\begin{equation}
\begin{aligned}{}
&\sum_{j=1}^N x^j_{s^k}\frac{\delta\Psi}{\delta x^j(s)}=0,\ \ k=1,\ldots,d,\\
&\sum_{j=1}^N-h^2\frac{\delta^2\Psi}{\delta x^j(s)^2}=
\sum_{1\le j_1<\ldots<j_d\le N}\left(\frac{\partial(x^{j_1},\ldots,x^{j_d})}
{\partial(s^1,\ldots,s^d)}\right)^2\Psi.
\end{aligned}
\end{equation}
This is a system of equations for a ``functional'' $\Psi$ of smooth functions $x(s)=(x^j(s))$, $j=1,\ldots,N$,
$s=(s^1,\ldots,s^d)$, $N>d$.

For formal derivation of equations (12), (13) from the generalized Hamilton--Jacobi equation
for the corresponding $(d+1)$-dimensional
variational problems, see [8].

\subsection{Two definitions of solutions of functional differential equations} 1) Let $V$ be the Schwartz
space of functions $\varphi(s)$, and $\Sc'$ be the corresponding space of distribution functionals (see \S2).
The space $\Sc'$ has an action of the Heisenberg Lie algebra consisting of pairs of functions
$(\varphi(s),\pi(s))\in V\oplus V'$,
with the canonical commutation relations
\begin{equation}
[\varphi_1,\varphi_2]=[\pi_1,\pi_2]=0,\ \ [\pi,\varphi]=ih\int\pi(s)\varphi(s)ds.
\end{equation}
By continuity, one can extend this action to an action of the algebra of functional differential operators
$S(V\oplus V')$ --- the topological symmetric algebra of the space $V\oplus V'$, consisting of
continuous polynomial functionals (symbols) $H=H(\varphi,\pi)$, with the $*$-product
\begin{equation}
\begin{aligned}{}
&H^{(1)}*H^{(2)}(\varphi,\pi)=\exp ih\int\frac{\delta}{\delta\pi_1(s)}\frac{\delta}{\delta\varphi_2(s)}\,ds\\
&\times H^{(1)}(\varphi_1,\pi_1)H^{(2)}(\varphi_2,\pi_2)|_{\varphi_1=\varphi_2=\varphi,\pi_1=\pi_2=\pi}.
\end{aligned}
\end{equation}
Let us denote this action by
$$
\Psi\in\Sc'\mapsto\widehat H\Psi\in\Sc'.
$$

\medskip

{\bf Definition.} Let us call by a {\it solution} of a system of differential equations
$(\widehat D_\alpha\Psi=0)$, $D_\alpha\in S(V'\oplus V')$,
in the space $\Sc'$, a distribution functional $\Psi\in\Sc'$ such that for any regularization
$D_{\alpha,\e}\in S(V\oplus V')$,
$\e>0$ of symbols $D_\alpha$, i.~e. $D_{\alpha,\e}\to D_\alpha$ as $\e\to0$, we have $\widehat D_{\alpha,\e}\Psi\to0$
in the space $\Sc'$ as $\e\to0$.
\medskip

This way we obtain definition of a solution $\Psi(t)\in\Sc'$ of the Schr\"odinger equation (11), and a
solution $\Psi(x(\cdot))\in\Sc'$,
depending on function $x(s)$, of the relativistically invariant generalization (12).
\medskip

2) One can also take as $V$ the space of functions $(x^\mu(s)$, $\varphi(s))$ (in the case of equation (12))
or the space of functions $(x^j(s))$ (in the case of quantum Plato problem (13)), and give, in exactly the same way,
definition of a solution $\Psi\in\Sc'$ of these equations, now not depending on additional parameters.
In this definition the space-time variables and the field variables participate in equal rights.

\subsection{Examples of solutions}

Here we can say only that traditional quantization of free scalar field in the Fock space fits into
our scheme, and yields solutions $\Psi(t)\in\Sc'$ of the Schr\"odinger equation (11) with
\begin{equation}
U(t,\x,\varphi)=m^2\varphi^2/2.
\end{equation}

Indeed, let us define the point $Z$ of the Siegel upper half-plane as the Lagrangian subspace
$L_0$ which is the image of the Schwartz space $V_\CC$ under the linear map $(Z_0,W_0):V_\CC\to V_\CC\oplus V'_\CC$,
where $W_0:V_\CC\to V'_\CC$ is the standard embedding, and $Z_0:V_\CC\to V_\CC$ is the operator $i\sqrt{-\Delta+m^2}$;
$\Delta$ is the Laplace operator in $\R^d$. It is easy to see that the point $Z$ is a fixed point of the
symplectic transform $V\oplus V\to V\oplus V$ corresponding to the (irregular) Hamiltonian
\begin{equation}
H(\varphi,\pi)=\int\frac12\left(\pi(\x)^2+
\sum_{j=1}^d\varphi_{x^j}(\x)^2+m^2\varphi(\x)^2\right)d\x.
\end{equation}
The point $Z$ determines, uniquely up to a number factor, the distribution functional $\Psi_Z\in\Sc'$ given by the
formula
\begin{equation}
\Psi_Z(\Phi)=\Phi((Z,0)),\ \ \Phi\in\Sc.
\end{equation}
This easily implies that $\Psi(t)=\Psi_Z$ is a stationary solution of the Schr\"odinger equation (11) in the sense
of the definition from the previous Subsection. Applying many times to $\Psi_Z$ elements
of the Lie algebra of shifts $V\subset V\oplus V'$, we obtain other, already non-stationary, solutions.

In other words, a dense subspace of the Fock space is embedded into the space $\Sc'$, and on this subspace
solutions of the Schr\"odinger equation (11) are given by usual formulas, as in quantization of free field.

It would be interesting to study the problem of finding solutions of equations (11--13) in the space $\Sc'$
different from the solutions given above. Besides that, if there are no new exact solutions, there can exist
asymptotic as $h\to0$ solutions. This would mean that one should generalize to infinite dimensions not the
Gauss--Fourier transform (3),
but the asymptotics of the integral (2) up to $o(h^\infty)$.
However, this investigation is beyond the scope of the present paper.

It would be also interesting to study formal perturbation theory of free field in the space $\Sc'$.

\end{document}